\documentclass[10pt]{elsarticle}
\usepackage{color}
\usepackage{slashed}
\usepackage{amsmath}
\usepackage{amssymb}
\usepackage{graphicx}
\usepackage{braket}
\newcommand{\beqn}{\begin{eqnarray}}
\newcommand{\eeqn}{\end{eqnarray}}

\definecolor{purple}{rgb}{0.8,0,0.6}

\newcommand{\revisionZ}[1]{{#1}}
\newcommand{\revisionS}[1]{{#1}}

\newcommand{\revisionZZ}[1]{{#1}}

\begin{document}

\title{Wigner - Weyl formalism and the propagator of Wilson fermions in the presence of varying external electromagnetic field}

\author[ARIEL]{M.Suleymanov}
\author[ARIEL]{M.A.~Zubkov
\footnote{
Corresponding author,
e-mail: zubkov@itep.ru}\footnote{on leave of absence from Institute for Theoretical and Experimental Physics, B. Cheremushkinskaya 25, Moscow, 117259, Russia} }

\address[ARIEL]{Physics Department, Ariel University, Ariel 40700, Israel}

\begin{abstract}
We develop  Wigner - Weyl formalism for the lattice models. For the definiteness we consider Wilson fermions in the presence of  $U(1)$ gauge field. The given technique reduces calculation of the two point fermionic Green function to solution of the Groenewold equation. It relates  Wigner transformation of the Green function with the Weyl symbol $Q_W$ of Wilson Dirac operator. We derive the simple expression for $Q_W$  in the presence of varying external $U(1)$ gauge field.  Next, we solve the Groenewold equation to all orders in powers of the derivatives of $Q_W$. Thus the given technique allows to calculate the fermion propagator in the lattice model with Wilson fermions in the presence of arbitrary background electromagnetic field. The generalization of this method to the other lattice models is straightforward.
\end{abstract}

\maketitle

\section{Introduction}

The Wigner - Weyl formalism was originally proposed for the reformulation of ordinary quantum mechanics. Later it was also  adopted in some form both for the quantum field theory and for the condensed matter physics theory. This formalism in its original form has been developed by H. Groenewold \cite{1} and J. Moyal \cite{2}.  It is based on the notions of the Weyl symbol of operator and the Wigner distribution function. The authors of \cite{1,2} used the ideas developed earlier by  H. Weyl \cite{3} and E. Wigner \cite{4}. The Wigner  - Weyl formalism in quantum mechanics is often referred to as the
phase space formulation. It is defined in phase space that is composed of both coordinates and momenta. At the same time the conventional formulation uses either coordinate space or  momentum space representations. In the phase space formulation the quantum state is described by the Wigner distribution (instead of a wave function), while the operator product is replaced by the Moyal product of functions defined in phase space.

The phase space formulation of quantum mechanics reduces the operator formulation in coordinate or momentum space to the formulation that deals with the ordinary functions of coordinates  and momenta \cite{5,berezin}.  There are a lot of various applications of the phase space formulation of quantum mechanics (for the review see, for example,  \cite{6,7}). At the present moment there exist the alternative versions of this formulation, in which the main notion - the distribution function - is defined differently  \cite{8,9}.  The most popular version mentioned above operates with the Wigner distribution function $W(x, p)$ \cite{4}. Among the alternative versions we mention those discussed in  \cite{10,11,12,13,14}.

In the one - dimensional case the quantum mechanical Wigner distribution function $W(x,p)$  depends on the coordinate $x$ and on momentum $p$. Distribution $W(x,p)$ determines the probability that the coordinate $x$ belongs to the interval $[a,b]$:
$${P} [a\leq x\leq b]=\frac{1}{2\pi}\int _{a}^{b}\int _{-\infty }^{\infty }W(x,p)\,dp\,dx
$$
In ordinary quantum mechanics each observable is represented by operator. Weyl symbol $A_W(x, p)$ of operator $\hat A$ is the function in phase space such that the expectation value of the observable with respect to the given distribution $W(x,p)$ is \cite{2,15}
$$
{ \langle {\hat {A}}\rangle =\frac{1}{2\pi}\int A_W(x,p)W(x,p)\,dp\,dx.}
$$
The Wigner function may be expressed through the ordinary wave function $\psi(x)$ as follows
$$
W(x,p) = \int dy \, e^{-ipy} \psi^*(x+y/2) \psi(x-y/2)
$$
Time evolution of the Wigner function is governed by the following equation
$$
i \partial_t W(x,p,t) = H(x,p)\star W(x,p,t)-W(x,p,t)\star H(x,p)
$$
Here the star (or Moyal) product of the two functions $f$ and $g$ is defined as
$${f\star g=f\,\exp {\left({\frac {i}{2}}({\overleftarrow {\partial }}_{x}{\overrightarrow {\partial }}_{p}-{\overleftarrow {\partial }}_{p}{\overrightarrow {\partial }}_{x})\right)}\,g} $$
The left arrow above the derivative means that it acts on $f$ while the right arrow means that it acts on $g$.
By $H(x,p)$ the Weyl symbol of the Hamiltonian is denoted. In general case the definition of the Weyl symbol of  operator $\hat{A}$ is:
$$
A_W(x,p) = \int dy \, e^{-ipy} \langle x+y/2| \hat{A} | x-y/2\rangle
$$

As it was mentioned above, the Wigner - Weyl formalism has also been extended, at least partially, to the quantum field theory. Namely, one may consider the analogue of the Wigner distribution function
$$
W_{\Gamma}(x,p) = \int dy \, e^{-ipy} \langle \bar{\psi}(x+y/2) \Gamma \psi(x-y/2)\rangle
$$
Here $\psi$ is a certain fermionic  operator while $\Gamma$ is an appropriate matrix in spinor and flavor spaces. (Obviously, the similar construction may be introduced also for bosons.) In particular, in this way the analogue of the Wigner distribution function has been defined in QCD (see, for example, \cite{QCDW,QCDW2} and references therein). Besides, the similar notion of the Wigner distribution has been implemented in the framework of the quantum kinetic theory based on the field theory \cite{KTW,KTW2}.  It was also used in certain noncommutative field theories \cite{NCW,NCW2}. Applications of the Wigner function formalism to cosmology were discussed in \cite{CSW}. Another applications of this formalism may be found, for example, in \cite{WW}, see also \cite{Berry}.

Recently the Wigner - Weyl formalism has been adopted to the investigation of anomalous transport in the quantum field theory \cite{ZW1,ZW2,ZW3,ZW4,ZW5,ZW6}. The main advantage of this latter approach is that the response of the nondissipative components of both vector and axial currents to the external field strength are expressed through the topological invariants in momentum space. These quantities are not changed if we deform the system smoothly. Thus we are able to calculate the corresponding conductivities within the simple noninteracting systems connected continuously with the more complicated interacting ones. And this gives the  values of the  conductivities for the given complicated systems. In particular, in this way the absence of the equilibrium chiral magnetic effect \cite{CME} in the field theory has been proved \cite{ZW5} while the anomalous quantum Hall effect has been described rigorously both for the Weyl semimetals and for certain $2D$ and $3D$ topological insulators \cite{ZW6}. Besides, the chiral separation effect (proposed in \cite{ CSE}) was rigorously re - derived \cite{ZW3} and the alternative derivation of chiral anomaly in lattice regularized quantum field theory has been given \cite{ZW1}. The fermion zero modes on vortices in the color superconductor phase of QCD have been considered using this technique in \cite{ZW4}.  In addition, the scale magnetic effect proposed in \cite{SME} has been considered in the framework of this formalism \cite{ZW2}.

It is worth mentioning that previously the momentum space topological invariants expressed in terms of the Green functions were used mainly in the condensed matter physics theory (for the review see
\cite{HasanKane2010,Xiao-LiangQi2011,Volovik2011,Volovik2007,Volovik2010}). The momentum space topological invariants protect gapless fermions on the boundaries of topological insulators \cite{Gurarie2011,EssinGurarie2011} and protect the bulk gapless fermions in Dirac and Weyl semi - metals \cite{Volovik2003,VolovikSemimetal}. The gapless fermions associated with topological defects and textures existing in the fermionic superfluids are also described by momentum space topology \cite{Volovik2016}. In the context of relativistic quantum field theory momentum space topology has been discussed, for example, in \cite{NielsenNinomiya1981,So1985,IshikawaMatsuyama1986,Kaplan1992,Golterman1993,Volovik2003,Horava2005,Creutz2008,Kaplan2011}. In \cite{VZ2012} the lattice regularization of QFT with Wilson fermions was discussed from the point of view of the momentum space topology. Appearance of the massless fermions at the intermediate values of bare mass parameter is related to the jump of the introduced momentum space topological invariant.

The version of the Wigner - Weyl formalism of \cite{ZW1,ZW3,ZW5,ZW6} deals with the lattice regularized quantum field theory. Here the two - point fermion Green function ${\cal G}({\bf p},{\bf q})$  in momentum space $\cal M$  (${\bf p}, {\bf q} \in {\cal M}$) has been considered. It obeys equation
$$
{\hat Q}({\bf p}-A(i {\partial_{\bf p}}))){\cal G}({\bf p},{\bf q}) = \delta({\bf p} - {\bf q})
$$
where ${\hat Q}({\bf p})$ is the lattice Dirac operator written in momentum space while $ A({\bf x})$ is external electromagnetic field. Its appearance in momentum representation is given by the pseudo - differential operator $ A(i {\partial_{\bf p}})$ realizing the so - called Peierls substitution. Notice, that in this formalism the imaginary time of the Euclidian formulation of the quantum field theory gives one of the components of the (discrete) coordinate space while the corresponding momentum is the fourth component of momentum vector that belongs to $\cal M$.

Wigner transformation of the two point Green function plays the role similar to the Wigner distribution in quantum mechanical formulation of \cite{1} and \cite{2}. It may be defined in continuous momentum space for the lattice model in the way similar to the Wigner transformation of continuous coordinate space:
\begin{equation} \begin{aligned}
{\cal G}_W({\bf x},p)\equiv\int dq e^{i{\bf x} {\bf q}} {\cal G}({{\bf p}+{\bf q}/2}, {{\bf p}-{\bf q}/2})\label{GWx0}
\end{aligned} \end{equation}
It obeys the Groenewold equation
\begin{equation} \begin{aligned}
{\cal G}_W({\bf x},{\bf p}) \star Q_W({\bf x},{\bf p}) = 1 \label{Geq0}
\end{aligned} \end{equation}
with the same star operation as above extended to the $D$ - dimensional vectors of coordinates ${\bf x}$ and momentum $\bf p$, while $Q_W$ is the Weyl symbol of operator $ {\hat Q}({\bf p}-A(i {\partial_{\bf p}}))$.
(Compare Eq. (\ref{Geq0}) with Eq. (4.38) of \cite{1} and Eq. (7.10) of \cite{2}, where the Weyl symbol of the commutator of two operators is represented.)

In the present paper we proceed with the development of the approach of
\cite{ZW1,ZW3,ZW5,ZW6}. For the definiteness we restrict ourselves with the lattice model with Wilson fermions in the presence of arbitrary external electromagnetic field. The obtained results may easily be extended to the other lattice models as well as to various tight - binding models of solid state physics. First of all we derive the precise expression for the Weyl symbol of the Wilson Dirac operator. Next, we present the complete iterative solution of the Groenewold equation Eq. (\ref{Geq0}), which allows to calculate the fermion propagator in the background of arbitrary external electromagnetic field.

The paper is organized as follows. We start in Sect. \ref{Sect1} from the pedagogical introduction to the conventional notions of Wigner transformation, Weyl symbols of operators and Moyal product. In Sect. \ref{Sect2} we proceed with the summary of the momentum space formulation of \cite{ZW1,ZW3,ZW5,ZW6}. In Sect. \ref{Sect3} we derive the Peierls substitution, that allows to calculate easily the Wilson Dirac operator in momentum space in the presence of arbitrary electromagnetic field. In Sect. \ref{Sect4} we present our results on the Weyl symbol of Wilson Dirac operator. In Sect. \ref{Sect5} we propose the procedure of the iterative solution of the Groenewold equation, which allows to calculate completely the  propagator of Wilson fermions in the presence of external electromagnetic field. In Sect \ref{Sect6} we end with the conclusions.

\section{Wigner transformation  in continuous coordinate space and continuous momentum space}

\label{Sect1}

In this section we briefly review  the technique of  Wigner transformation applied to quantum mechanics  defined in infinite continuous coordinate space.

\subsection{Weyl symbol of operator}

We start from the definition of the average of  operator $\hat{A}$  with respect to quantum state $\Psi$
\begin{equation}
\bra{\Psi} \hat{A} \ket{\Psi}=\int dx dy \braket{\Psi|x}  \bra{x} \hat{A} \ket{y}  \braket{y|\Psi}
\label{Z} \end{equation}
Here by $x, y$ we denote the continuous coordinates. For simplicity we consider the case of one - dimensional space $R^1$. The generalization of our expressions to the case of $D$ - dimensional space $R^D$  is straightforward.  Let us change the coordinates:
\begin{equation}\begin{aligned}
x=u+v/2 \,\,\,\,\,\,\,\,\, y=u-v/2
\label{Z} \end{aligned} \end{equation}
Then
\begin{equation}\begin{aligned}
dxdy=\frac{\partial(x,y)}{\partial(u,v)}dudv=
\left |
\begin{matrix}
\frac {\partial x}{\partial u} & \frac {\partial y}{\partial u}\\
\frac {\partial x}{\partial v} & \frac {\partial y}{\partial v}
\end{matrix}
\right |
dudv=-dudv
\label{Z} \end{aligned} \end{equation}
This gives
\begin{equation} \begin{aligned}
&\bra{\Psi} \hat{A} \ket{\Psi}
=-\int dx dy \bra{x+y/2} \hat{A} \ket{x-y/2} \braket{\Psi|x+y/2} \braket{x-y/2|\Psi}=\\
-&\int dx dy dz \bra{x+y/2} \hat{A} \ket{x-y/2} \delta(z-y) \braket{\Psi|x+z/2} \braket{x-z/2|\Psi}=\\
-&\int dx dy dz dp \bra{x+y/2} \hat{A} \ket{x-y/2} \frac{e^{-ip(z-y)}}{2\pi} \braket{x-z/2|\Psi} \braket{\Psi|x+z/2}= \\
&\int \frac{dx dp}{2\pi} dy e^{ipy} \bra{x+y/2} \hat{A} \ket{x-y/2} dz e^{ipz} \braket{x+z/2|\Psi} \braket{\Psi|x-z/2}
\label{Z}\end{aligned} \end{equation}
The {\bf Weyl symbol of operator} $\hat{A}$  is  defined as follows
\begin{equation} \begin{aligned}
A_W(x,p)\equiv\int  dy e^{-ipy} \bra{x+y/2} \hat{A} \ket{x-y/2}
\label{Z}\end{aligned} \end{equation}
or, in terms of the matrix elements in momentum space:
\begin{equation} \begin{aligned}
A_W(x,p)\equiv\int  dq e^{iqx} \bra{p+q/2} \hat{A} \ket{p-q/2}
\label{Z}\end{aligned} \end{equation}
Here
$$
\langle {x} \ket{p} = \frac{1}{\sqrt{2\pi}}e^{i p x}
$$
Wigner function for the state $\Psi$ is defined as the Weyl symbol of the corresponding density operator:
\begin{equation} \begin{aligned}
W(x,p)=\rho_W(x,p)=\int dy  e^{-ipy} \braket{x+y/2|\Psi} \braket{\Psi|x-y/2}
\label{Z}\end{aligned} \end{equation}
In momentum space we have:
\begin{equation} \begin{aligned}
W(x,p)=\rho_W(x,p)=\int dq  e^{iqx} \braket{p+q/2|\Psi} \braket{\Psi|p-q/2}
\label{Z}\end{aligned} \end{equation}
Hence,

\begin{equation} \begin{aligned}
\bra{\Psi} \hat{A} \ket{\Psi}=\int \frac{dxdp}{2\pi} A_W(x,p)W(x,p)
\label{Z}\end{aligned} \end{equation}

Wigner function for the mixed state with the density matrix
$$
\hat{\rho} =\sum_i \ket{\Psi_i}\rho_i\bra{\Psi_i}
$$
may be defined as
\begin{equation} \begin{aligned}
W(x,p)=\rho_W(x,p)=\int dq  e^{iqx} \sum_i\braket{p+q/2|\Psi_i} \rho_i\braket{\Psi_i|p-q/2}
\label{Z}\end{aligned} \end{equation}
The vacuum average of operator $\hat{A}$ with respect to this mixed state may be written in the similar way
\begin{equation} \begin{aligned}
\langle \hat{A} \rangle =\int \frac{dxdp}{2\pi} A_W(x,p)W(x,p)
\label{Z}\end{aligned} \end{equation}

\subsection{Moyal product}
The Weyl symbol of the product of two operators is given by
\begin{equation}\begin{aligned}
(AB)_W(x,p)=&\int dy e^{-ipy} \Bra{x+y/2} \hat{A} \hat{B} \ket{x-y/2}=\\
&\int dy dz e^{-ipy} \bra{x+y/2} \hat{A} \ket{z}\bra{z}\hat{B} \ket{x-y/2}
\label{Z}\end{aligned}\end{equation}
The transformation of variables
\begin{equation}\begin{aligned}
y=u+v \,\,\,\,\,\,\,\,\, z=x-u/2+v/2
\label{Z} \end{aligned} \end{equation}
gives
\begin{equation}\begin{aligned}
dydz=\frac{\partial(y,z)}{\partial(u,v)}dudv=
\left |
\begin{matrix}
\frac {\partial y}{\partial u} & \frac {\partial z}{\partial u}\\
\frac {\partial y}{\partial v} & \frac {\partial z}{\partial v}
\end{matrix}
\right |
dudv=dudv
\label{Z} \end{aligned} \end{equation}
and
\begin{equation}\begin{aligned}
&(AB)_W(x,p)=\\
& \int du dv e^{-ip(u+v)}
\Bra{x+\frac{u}{2}+\frac{v}{2}} \hat{A} \Ket{x-\frac{u}{2}+\frac{v}{2}}
\Bra{x-\frac{u}{2}+\frac{v}{2}}\hat{B} \Ket{x-\frac{u}{2}-\frac{v}{2}}=\\
&\left [ \int du e^{-ipu}
\Bra{x+\frac{u}{2}} \hat{A} \Ket{x-\frac{u}{2}} \right]
e^{\frac{v}{2}\overleftarrow{\partial_x}-\frac{u}{2}\overrightarrow{\partial_x}}
\left[ \int dve^{-ipv}
\Bra{x+\frac{v}{2}}\hat{B} \Ket{x-\frac{v}{2}} \right]=\\
&\left [ \int du e^{-ipu}
\Bra{x+\frac{u}{2}} \hat{A} \Ket{x-\frac{u}{2}} \right ]
e^{\frac{i}{2} \left( \overleftarrow{\partial_x}\overrightarrow{\partial_p}-\overleftarrow{\partial_p}\overrightarrow{\partial_x}\right )}
\left [ \int dve^{-ipv}
\Bra{x+\frac{v}{2}}\hat{B} \Ket{x-\frac{v}{2}} \right ]
\label{Z}\end{aligned}\end{equation}
{\bf The Moyal product} of the Weyl symbols of the two operators $\hat A$ and $\hat B$ is defined as follows
\begin{equation}\begin{aligned}
& A_W(x,p) \star B_W(x,p) \equiv (AB)_W(x,p)=A_W(x,p) e^{\frac{i}{2} \left( \overleftarrow{\partial_x}\overrightarrow{\partial_p}-\overleftarrow{\partial_p}\overrightarrow{\partial_x}\right )} B_W(x,p)
\label{Z}\end{aligned}\end{equation}

\section{Wigner transformation in discrete coordinate space and compact momentum space}
\label{Sect2}

In this section we discuss the generalization of the above mentioned constructions to the case, when coordinate space is discrete and infinite, and, therefore, momentum space is continuous but compact.
\subsection{Weyl symbol of operator}
Average of an operator with respect to state $\Psi$ is defined as
\begin{equation}
\bra{\Psi} \hat{A} \ket{\Psi}=\int dp dq \braket{\Psi|p}  \bra{p} \hat{A} \ket{q}  \braket{q|\Psi}
\label{Z} \end{equation}
Here the integral is over the compact momentum space. The integration measure is defined in such a way, that $$
\int dp =2\pi
$$
while
$$
\langle x_n|p\rangle = \frac{1}{\sqrt{2\pi}} e^{i p x_n}
$$
for any lattice point $x_n$.
Again, we restrict ourselves by the one dimensional case. The generalization to the multidimensional systems is straightforward.  Then
\begin{equation} \begin{aligned}
\bra{\Psi} \hat{A} \ket{\Psi}
=-&\int dp dq \bra{p+q/2} \hat{A} \ket{p-q/2} \braket{\Psi|p+q/2} \braket{p-q/2|\Psi}\\=
-&\int dp dq dk \bra{p+q/2} \hat{A} \ket{p-q/2} \delta(k-q) \braket{\Psi|p+k/2} \braket{p-k/2|\Psi}\\=
-&\sum_{x_n} \int dp dq dk \bra{p+q/2} \hat{A} \ket{p-q/2} \frac{e^{-ix_n(k-q)}}{2\pi} \braket{p-k/2|\Psi} \braket{\Psi|p+k/2} \\=
&\sum_{x_n} \int \frac{dp}{2\pi} dq e^{ix_nq} \bra{p+q/2} \hat{A} \ket{p-q/2} dk e^{ix_nk} \braket{p+k/2|\Psi} \braket{\Psi|p-k/2}
\label{Z}\end{aligned} \end{equation}
The Weyl symbol of operator $\hat{A}$ is then defined as follows
\begin{equation} \begin{aligned}
A_W(x_n,p)\equiv\int dq e^{ix_nq} \bra{p+q/2} \hat{A} \ket{p-q/2}\label{AWx}
\end{aligned} \end{equation}
And the Wigner function, which is the Weyl symbol of density operator, is
\begin{equation} \begin{aligned}
W(x_n,p)=\int dq e^{ix_nq} \braket{p+q/2|\Psi} \braket{\Psi|p-q/2}\label{Wx}
\end{aligned} \end{equation}
hence
\begin{equation}\begin{aligned}
\bra{\Psi} \hat{A} \ket{\Psi}= \sum_{x_n} \int \frac{dp}{2\pi} A_W(x_n,p)W(x_n,p)
\label{Z}\end{aligned} \end{equation}

\subsection{Moyal product}
Let us consider the Weyl symbol of the product of two operators $\hat A$ and $\hat B$:
\begin{equation}\begin{aligned}
&(AB)_W(x_n,p)=\\
& \int dq dk e^{ix_n(q+k)}
\Bra{p+\frac{q}{2}+\frac{k}{2}} \hat{A} \Ket{p-\frac{q}{2}+\frac{k}{2}}
\Bra{p-\frac{q}{2}+\frac{k}{2}}\hat{B} \Ket{p-\frac{q}{2}-\frac{k}{2}}=\\
&\left [ \int dq e^{ix_nq}
\Bra{p+\frac{q}{2}} \hat{A} \Ket{p-\frac{q}{2}} \right]
e^{\frac{k}{2}\overleftarrow{\partial_p}-\frac{q}{2}\overrightarrow{\partial_p}}
\left[ \int dk e^{ix_nk}
\Bra{p+\frac{k}{2}}\hat{B} \Ket{p-\frac{k}{2}} \right]=\\
&\left [ \int dq e^{ix_nq}
\Bra{p+\frac{q}{2}} \hat{A} \Ket{p-\frac{q}{2}} \right ]
e^{\frac{i}{2} \left(- \overleftarrow{\partial_p}\overrightarrow{\partial}_{x_n}+\overleftarrow{\partial}_{x_n}\overrightarrow{\partial_p}\right )}
\left [ \int dk e^{ix_nk}
\Bra{p+\frac{k}{2}}\hat{B} \Ket{p-\frac{k}{2}} \right ]
\label{Z}\end{aligned}\end{equation}
Hence, the {\bf Moyal product} may be defined as in the case of continuous space
\begin{equation}\begin{aligned}
&(AB)_W(x_n,p)=
A_W(x_n,p)
e^{\frac{i}{2} \left( \overleftarrow{\partial}_{x_n}\overrightarrow{\partial_p}-\overleftarrow{\partial_p}\overrightarrow{\partial}_{x_n}\right )}
B_W(x_n,p)
\label{ZAB}\end{aligned}\end{equation}
Notice, that Eqs. (\ref{AWx}) and (\ref{Wx}) define the Weyl symbol of operator $\hat{A}$ and the Wigner function as the functions of the continuous variable $x_n$ although the original coordinate space is discrete. This formal definition allows to use Eq. (\ref{ZAB}) with the derivative with respect to $x_n$.

\revisionZ{As it was mentioned above, all considered expressions may be easily extended to the case of the D - dimensional system with the compact momentum space of arbitrary form.  Then we assume that the integration measure over momentum space is normalized in such a way, that
$$
\int dp = (2\pi)^D
$$
while the states $\ket{x_n}$ are defined in such a way that $\bra{x_n}x_m\rangle = \delta_{nm}$.
With this normalization the definitions of the Weyl symbol of operator and the Wigner function as well as the definition of Moyal product remain the same. At the same time for the average of operator $\hat A$ with respect to the given quantum state  we have
\begin{equation}\begin{aligned}
\bra{\Psi} \hat{A} \ket{\Psi}= \sum_{x_n} \int \frac{dp}{(2\pi)^D} A_W(x_n,p)W(x_n,p)
\label{Z}\end{aligned} \end{equation}
For the mixed states given by the density matrix $\hat{\rho} = \sum_i\ket{ \Psi_i} \rho_i\bra{\Psi_i} $ we have the Wigner function of the form
\begin{equation} \begin{aligned}
W(x,p)=\rho_W(x,p)=\int dq  e^{iqx} \sum_i\braket{p+q/2|\Psi_i} \rho_i\braket{\Psi_i|p-q/2}
\label{Z}\end{aligned} \end{equation}
Again, the vacuum average of operator $\hat{A}$ with respect to this mixed state may be written as
\begin{equation} \begin{aligned}
\langle \hat{A} \rangle =\int \frac{dxdp}{(2\pi)^D} A_W(x,p)W(x,p)
\label{Z}\end{aligned} \end{equation}}

\section{Wilson fermions and Peierls substitution }

\label{Sect3}

\subsection{Free Wilson fermions}

Let us consider the partition function for the lattice model with Wilson fermions written in the discrete coordinate space:
\begin{equation}
Z=\int D\bar{\Psi}D\Psi exp\left(-\sum_{{\bf r}_n,{\bf r}_m}\bar\Psi({\bf r}_m)\left(-i\mathcal{D}_{{\bf r}_n,{\bf r}_m}\right)\Psi({\bf r}_n)\right)
\label{Z} \end{equation}
Here
\begin{equation}
\mathcal{D}_{\bf x,y}=-\frac{1}{2}\sum_i \left[ (1+\gamma^i)\delta_{{\bf x+e}_i,{\bf y}}+
(1-\gamma^i)\delta_{{\bf x-e}_i,{\bf y}}\right]U_{{\bf x},{\bf y}}+
(m^{(0)}+4)\delta_{{\bf x},{\bf y}}
\label{D} \end{equation}
$\gamma^k$ are the Dirac gamma matrices, $m^{(0)}$ is the parameter that has the meaning of mass.

{Wilson fermions in momentum space correspond to the propagator of the form
$$
{\cal G}({\bf p}) = \hat{Q}^{-1}({\bf p})
$$
Here the operator $\hat Q$ has the form
\begin{equation} \begin{aligned}
\hat{Q}({\bf p}) =\sum_{k=1,2,3,4} \gamma^k g_k ({\bf p})-im({\bf p})
\label{WF0}\end{aligned} \end{equation}
where
\begin{equation} \begin{aligned}
g_k({\bf p})=\sin( p_k) \quad\quad m({\bf p})=
m^{(0)}+\sum_{\nu=1}^4 (1-\cos(p_\nu))
\label{Z}\end{aligned} \end{equation}
The same system represents the cubic tight - binding model of certain solid state systems if momentum component $p_4$ is rescaled accordingly in order to represent the discretization of the imaginary time. The model may be extended in several ways to reproduce the tight - binding models for various real materials.

Below we will show that coupling of this system to the electromagnetic field ${\bf A}(x)$ results in the so - called Peierls substitution $\hat { \bf p} \rightarrow \hat {\bf p}-{\bf A}(i\partial_{\bf p})$.  In momentum space the electromagnetic field appears as the pseudo - differential operator, in which the dependence of $\bf A$  on coordinate is substituted by the dependence on the operator  $i\partial_{\bf p}$. }

\subsection{Wilson fermions in the presence of external Abelian gauge field}

Let us consider the partition function for the lattice model with Wilson fermions in the presence of the gauge field written in discrete coordinate space:
\begin{equation}
Z=\int D\bar{\Psi}D\Psi exp\left(-\sum_{{\bf r}_n,{\bf r}_m}\bar\Psi({\bf r}_m)\left(-i\mathcal{D}_{{\bf r}_n,{\bf r}_m}\right)\Psi({\bf r}_n)\right)
\label{Z} \end{equation}
Here
\begin{equation}
\mathcal{D}_{\bf x,y}=-\frac{1}{2}\sum_i \left[ (1+\gamma^i)\delta_{{\bf x+e}_i,{\bf y}}+
(1-\gamma^i)\delta_{{\bf x-e}_i,{\bf y}}\right]U_{{\bf x},{\bf y}}+
(m^{(0)}+4)\delta_{{\bf x},{\bf y}}
\label{D} \end{equation}
$\gamma^k$ are the Dirac gamma matrices, $m^{(0)}$ is the parameter that has the meaning of mass, while
\begin{equation}
U_{\bf x,y}=Pe^{i\int_{\bf x}^{\bf y} d {\pmb \xi} {\bf A}({\pmb \xi})}
\label{D} \end{equation}
We restrict ourselves by the case of the $U(1)$  gauge field $\bf A$ and then this parallel transporter is given by
\begin{equation}
U_{\bf x,y}=e^{i\int_{\bf x}^{\bf y} d {\pmb \xi} {\bf A}({\pmb \xi})}
\label{D} \end{equation}

\subsection{Peierls substitution}
Let us denote
\begin{equation}\begin{aligned}
\mathcal{D}_{{\bf r}_n,{\bf r}_m}=-\frac{1}{2}\sum_i \left[ (1+\gamma^i)\delta_{{\bf r}_n+{\bf e}_i,{\bf r}_m}+
(1-\gamma^i)\delta_{{\bf r}_n-{\bf e}_i,{\bf r}_m}\right]U_{{\bf r}_n,{\bf r}_m}+
(m^{(0)}+4)\delta_{{\bf x},{\bf y}}
\label{D1} \end{aligned}\end{equation}
and
\begin{equation}
\begin{aligned}
I_1=\sum_{{\bf r}_n,{\bf r}_m}\bar \Psi({\bf r}_m)\left(-i\mathcal{D}_{{\bf r}_n,{\bf r}_m}\right)\Psi({\bf r}_n)
\label{3} \end{aligned} \end{equation}
Also we define
\begin{equation}
\Psi ({\bf r})=\int_{\mathcal M} \frac{d^D {\bf p}}{|\mathcal M|}e^{i{\bf rp}}\Psi ({\bf p})
\label{4} \end{equation}
where $|\mathcal M| = (2\pi)^D$ is the volume of momentum space.
\\
The above expression for $I_1$ contains the terms proportional to
\begin{equation} \begin{aligned}
&\sum_{{\bf r}_n,{\bf r}_m}
\bar \Psi({\bf r}_m) \delta_{{\bf r}_n\pm{\bf e}_i,{\bf r}_m}
e^{i\int_{{\bf r}_n}^{{\bf r}_m} d {\pmb \xi} {\bf A}({\pmb \xi})}
\Psi({\bf r}_n)=\\
&\sum_{{\bf r}_n}
\bar \Psi({\bf r}_n\pm{\bf e}_i)
e^{i\int_{{\bf r}_n}^{{\bf r}_n \pm {\bf e}_i} d {\pmb \xi} {\bf A}({\pmb \xi})}
\Psi({\bf r}_n)=\\
& \sum_{{\bf r}_n}
\left[\int_{\mathcal M} \frac{d^D {\bf p}}{|\mathcal M|}e^{-i({\bf r}_n \pm {\bf e}_i){\bf p}}\bar \Psi ({\bf p})\right]
e^{i\int_{{\bf r}_n}^{{\bf r}_n \pm {\bf e}_i} d {\pmb \xi} {\bf A}({\pmb \xi})}
\left[\int_{\mathcal M} \frac{d^D {\bf q}}{|\mathcal M|}e^{i{\bf r}_n{\bf q}}\Psi ({\bf q})\right]=\\
&\sum_{{\bf r}_n}
\int_{\mathcal M} \frac{d^D {\bf p}}{|\mathcal M|}\int_{\mathcal M} \frac{d^D {\bf q}}{|\mathcal M|}
\bar \Psi ({\bf p})e^{i{\bf r}_n{\bf  q}}
e^{i\int_{{\bf r}_n}^{{\bf r}_n \pm {\bf e}_i} d {\pmb \xi} {\bf A}({\pmb \xi})}
e^{-i({\bf r}_n\pm {\bf e}_i){\bf p}}
\Psi ({\bf q})=\\
&\sum_{{\bf r}_n}
\int_{\mathcal M} \frac{d^D {\bf p}}{|\mathcal M|}\int_{\mathcal M} \frac{d^D {\bf q}}{|\mathcal M|}
\bar \Psi ({\bf p}) e^{i{\bf r}_n{\bf  q}}
\exp\left[-i\int_0^{\mp {\bf e}_i} d \xi {\bf e}_i e^{i\xi {\bf e}_i {\bf p}} {\bf A} ({i\partial_{\bf p}}) e^{-i\xi {\bf e}_i {\bf p}}\right]
e^{-i({\bf r}_n \pm {\bf e}_i){\bf p}}
\Psi ({\bf q})=\\
&
\int_{\mathcal M} \frac{d^D {\bf p}}{|\mathcal M|}
\bar \Psi ({\bf p})
\exp\left[-i\int_0^{\mp 1} d \xi e^{i\xi p_i} A_i ({i\partial_{\bf p}})e^{-i\xi p_i}\right]
e^{\mp i{\bf e}_i{\bf p}}
\Psi ({\bf p})
\label{I1_}
\end{aligned} \end{equation}
Here we used Eq. (\ref{Lemma2}) from Appendix II with $x = {\bf r}_n \pm {\bf e}_i$. Now let us apply Eq. (\ref{Lemma1}) from the same Appendix. This gives
\begin{equation}\begin{aligned}
&\sum_{{\bf r}_n,{\bf r}_m}
\bar \Psi({\bf r}_m) \delta_{{\bf r}_n\pm{\bf e}_i,{\bf r}_m}
e^{i\int_{{\bf r}_n}^{{\bf r}_m} d {\pmb \xi} {\bf A}({\pmb \xi})}
\Psi({\bf r}_n)=\\
&
\int_{\mathcal M} \frac{d^D {\bf p}}{|\mathcal M|}
\bar \Psi ({\bf p})
e^{\mp i{\bf e}_i({\bf p}-A_i ({i\partial_{\bf p}})) }
\Psi ({\bf p})
\label{I1}
\end{aligned} \end{equation}
We come to the partition function of the form
\begin{equation}
Z=\int D\bar{\Psi}D\Psi exp\left(-\int \frac{d^D {\bf p}}{|{\cal M}|}\bar\Psi({\bf p})  Q({\bf p} - {\bf A}(i\partial_{\bf p}))\Psi({\bf p})\right)
\label{Z} \end{equation}
where $Q$ is given by Eq. (\ref{WF0}). One can see, that in momentum space the  application of the external Abelian gauge field results in the Peierls substitution
$$
{\bf p} \to {\bf p} - {\bf A}(i\partial_{\bf p})
$$

\section{Weyl symbol for the Wilson fermions}
\label{Sect4}

\subsection{Weyl symbols of the simplest operators}
Below we will be interested in the Weyl symbols of the operators of the form of $Q({\bf p} - {\bf A}(i\partial_{\bf p}) )$, where $A$ is given by Eq. (\ref{WF0}).
Let us start our consideration from the Weyl symbol for a general function of ${\bf p}$
\begin{equation} \begin{aligned}
(f({\bf p}))_W({\bf x,p})=&\int  d{\bf q} e^{i\bf qx} \bra{\bf p+q/2} f({\bf p}) \ket{\bf p-q/2} =\\
&\int  d{\bf q} e^{i\bf qx} f({\bf p-q/2}) \braket{\bf p+q/2|p-q/2} =\\
&\int  d{\bf q} e^{i\bf qx} f({\bf p-q/2})\delta({\bf q})=f({\bf p})
\label{Z}\end{aligned} \end{equation}
Weyl symbol for a general function of $i\partial_{\bf p}$ is
\begin{equation} \begin{aligned}
( g(i\partial_{\bf p}))_W({\bf x,p})=&\int  d{\bf q} e^{i\bf qx} \bra{\bf p+q/2} g(i\partial_{\bf p}) \ket{\bf p-q/2} =\\
&\int  d{\bf q} e^{i\bf qx} \bra{\bf p+q/2} g(i\partial_{\bf p}) \ket{\bf y} \braket{\bf y|p-q/2}  d{\bf y} =\\
&\int  d{\bf q} e^{i\bf qx} \bra{\bf p+q/2} g(i\partial_{\bf p}) \ket{\bf y}  \frac{e^{i\bf (p-q/2)y}}{(2\pi)^{3/2}} d{\bf y} =\\
&\int  d{\bf q} e^{i\bf qx} g({\bf y}) \braket{\bf p+q/2|y}   \frac{e^{i\bf (p-q/2)y}}{(2\pi)^{3/2}} d{\bf y} =\\
&\int  d{\bf q} e^{i\bf qx} g({\bf y})  \frac{e^{-i\bf (p+q/2)y}}{(2\pi)^{3/2}}  \frac{e^{i\bf (p-q/2)y}}{(2\pi)^{3/2}} d{\bf y} =\\
&\int  d{\bf q} \frac{e^{i\bf q(x+y)}}{(2\pi)^3}  g({\bf y})   d{\bf y} =g({\bf x})\\
\label{Z}\end{aligned} \end{equation}
Here we used the fact that $i\frac{\partial}{\partial p} \ket{y} = y \ket{y}$.
In these expressions  for the definiteness we restricted ourselves by the case of the three - dimensional momentum space.

\subsection{Weyl symbol of $\exp\big(i(\hat p_k-A_ke^{-{\pmb \omega} {\partial_{\bf p}}})\big)$ }

Operator $\hat Q$ for the Wilson fermions is composed of the functions of a single momentum component. Moreover, those functions are $sin$ and $cos$ that are given by the sum of exponents. Therefore, in order to find the Weyl symbol of $\hat Q$ it is enough to calculate the Weyl symbol of the exponent of the combination $p_k - A(p_1,...,p_D)$. We start our calculation from the consideration of
$F( {\bf p},i\partial_{\bf p})=
\exp\big(i(p_k-A_ke^{-{\pmb \omega} {\partial_{\bf p}}})\big)$

Let us calculate the commutator
 $[p_k ,A_k e^{-{\pmb \omega} {\partial_{\bf p}}}]$:
\begin{equation} \begin{aligned}
iA_k e^{-{\pmb \omega} {\partial_{\bf p}}}i p_k=
i(p_k-\omega_k)iA_k e^{-{\pmb \omega} {\partial_{\bf p}}}
\label{Z}\end{aligned} \end{equation}
Therefore,
\begin{equation}
[i\hat p_k, iA_k e^{-{\pmb \omega} {\partial_{\bf p}}}]=i\omega_k iAe^{-{\pmb \omega} {\partial_{\bf p}}}
\label{Z} \end{equation}
Using the special case of BCH formula where $[X,Y]=\alpha Y$, we obtain:
\begin{equation} \begin{aligned}
e^Xe^Y=e^{X+\frac{\alpha}{1-e^{-\alpha}}Y}
\label{Z}\end{aligned} \end{equation}
\revisionZZ{This expression is given, for example, in Eq. (1.5) of  \cite{Su}. The rather sophisticated derivation of this formula may be found, for example, in \cite{Wi}. We present our derivation of this formula in Appendix III.}
Changing variables to $Y'\equiv \frac{\alpha}{1-e^{-\alpha}}Y=\beta^{-1} Y$ ,
where $\beta^{-1} \equiv \frac{\alpha}{1-e^{-\alpha}}$, and  $[X,Y']=\alpha Y'$ , we obtain:
\begin{equation} \begin{aligned}
e^{X+Y'}=e^Xe^{\beta Y'}
\label{Z}\end{aligned} \end{equation}
In case of $\exp\big(i(\hat p_k-A_k e^{-{\pmb \omega} {\partial_{\bf p}}})\big)$ with $\alpha=i\omega_k$
we come to
\begin{equation} \begin{aligned}
\exp\big(i(\hat p_k-A_k e^{-{\pmb \omega} {\partial_{\bf p}}})\big)=
e^{i \hat p_k} \exp(-i\beta A_ke^{-{\pmb \omega} {\partial_{\bf p}}})
\label{Z}\end{aligned} \end{equation}
This decomposition allows to calculate the Weyl symbol of $F( {\bf p},i\partial_{\bf p})$:
\begin{equation} \begin{aligned}
&\Big[\exp\big(i(\hat p_k -A_k e^{-{\pmb \omega} {\partial_{\bf p}}})\big)\Big]_W=
\Big[e^{i\hat p_k} \exp(-i\beta A_ke^{-{\pmb \omega} {\partial_{\bf p}}})\Big]_W=\\
&\int  d{\bf q} e^{i\bf qx} \bra{\bf p+q/2}
e^{i\hat p_k} \exp(-i\beta A_ke^{-{\pmb \omega}{\partial_{\bf p}}})
 \ket{\bf p-q/2} \\
\label{Z}\end{aligned} \end{equation}
\revisionS{Next, we use that  $i\partial_p \ket{y}  = y \ket{y} $ and
$$
i\partial_p \ket{p-q/2} = \int dy i \partial_p \ket{y}\braket{y | p-q/2} = \frac{1}{(2\pi)^{3/2}}\int dy y \ket{y}e^{i y ( p-q/2)} = \frac{1}{(2\pi)^{3/2}}\int dy  \ket{y} (-i\partial_p ) e^{i y ( p-q/2)}
$$
Therefore,
\begin{equation} \begin{aligned}
&e^{i\hat p_k} \exp(-i\beta A_ke^{-{\pmb \omega}{\partial_{\bf p}}}) \ket{\bf p-q/2 }=
e^{i\hat p_k} \sum_{n=0}^\infty \frac{(-i\beta A_k)^n}{n!} e^{-n{\pmb \omega}{\partial_{\bf p}}} \ket{\bf p-q/2}=\\
&e^{i\hat p_k} \sum_{n=0}^\infty \frac{(-i\beta A_k)^n}{n!} \ket{{\bf p -q/2}+n{\pmb \omega}}=
\sum_{n=0}^\infty e^{i( p_k-q_k/2+n\omega_k)}\frac{(-i\beta A_k)^n}{n!} \ket{{\bf p-q/2}+n{\pmb \omega}}
\label{Z}\end{aligned} \end{equation}}
and come to
\begin{equation} \begin{aligned}
&\Big[\exp\big(i(\hat p_k -A_k e^{-{\pmb \omega} {\partial_{\bf p}}})\big)\Big]_W=\\
&\int  d{\bf q} e^{i\bf qx}
\sum_{n=0}^\infty e^{i(p_k-q_k/2+n\omega_k)}\frac{(-i\beta A_k)^n}{n!} \braket{{\bf p+q/2}|{\bf p-q/2}+n{\pmb \omega}}=\\
& \sum_{n=0}^\infty e^{i n{\pmb \omega} {\bf x}} e^{i(p_k+n\omega_k/2)}\frac{(-i\beta A_k)^n}{n!}=
e^{ip_k} \sum_{n=0}^\infty e^{ ni({\pmb \omega} {\bf x}+\omega_k/2)}\frac{(-i\beta A_k)^n}{n!}= \\
&\exp{\Big[i(p_k -\beta A_k e^{ i({\pmb \omega} {\bf x}+\omega_k/2)})\Big]}=
\exp{\Big[i\big(p_k - \frac{(1-e^{-i\omega_k})e^{i\omega_k/2}}{i\omega_k}A_k e^{ i{\pmb \omega} {\bf x}}\big)\Big]}\\
=&\revisionZ{\exp{\Big[i\big(p_k-\frac{\sin(\omega_k/2)}{\omega_k/2}A_k e^{ i{\pmb \omega} {\bf x}}\big)\Big]}}
\label{Z}\end{aligned} \end{equation}

\subsection{Weyl symbol of $F=
\exp\big(i(\hat p_\mu - \sum_{J=1}^N A_{J \mu}e^{ {\bf k}_J i\partial_{\bf p}})\big)$ }

The next step is the consideration of the function $F=
\exp\big(i(\hat p_\mu -  \sum_{J=1}^N A_{J \mu}e^{ {\bf k}_J i\partial_{\bf p}})\big)$, which appears in the series expansion for arbitrary function (Fourier or Laplace series).
As above, we will be using the following relation:
\begin{equation} \begin{aligned}
iA_{J\mu} e^{{\bf k}_J i\partial_{\bf p}} i\hat p_\mu=
i( \hat p_\mu+ik_{J\mu})iA_{J\mu} e^{{\bf k}_J i\partial_{\bf p}}
\label{Z}\end{aligned} \end{equation}
It gives
\begin{equation}
 [i\hat p_\mu, iA_{J\mu} e^{{\bf k}_J i\partial_{\bf p}}] =
k_{J\mu} iA_{J\mu} e^{{\bf k}_J i\partial_{\bf p}}
\label{Z} \end{equation}
Let us define the operator
\begin{equation}
\hat B_{J\mu} \equiv A_{J\mu} e^{{\bf k}_J i\partial_{\bf p}}
\label{Z}\end{equation}
It obeys the commutation relations
\begin{equation}
[\hat B_{J\mu}, \hat B_{I\mu}]=0
\label{Z}\end{equation}
and
\begin{equation}
 [i\hat p_\mu, i\hat B_{J\mu} ] =
k_{J\mu} i\hat B_{J\mu}
\label{Z} \end{equation}
Therefore, we have
\begin{equation}
 [i(\hat p_\mu-B_{J\mu}), iB_{I\mu} ] =k_{I\mu} iB_{I\mu}
\label{Z} \end{equation}
and
\begin{equation} \begin{aligned}
 \Big[i\Big(\hat p_\mu-\sum_{J} \hat B_{J\mu}\Big), i\hat B_{I\mu} \Big] =k_{I\mu} i\hat B_{I\mu}
\label{Z}\end{aligned} \end{equation}
As above we are able to use the special case of \revisionS{the BCH formula \cite{Su}} for $[X,Y]=\alpha Y$, $\beta=\frac{1-e^{-\alpha}}{\alpha}$
\begin{equation} \begin{aligned}
e^{X+Y}=e^Xe^{\beta Y}
\label{Z}\end{aligned} \end{equation}
 We come to the following representation:
\begin{equation} \begin{aligned}
&\exp\Big({i\hat p_\mu -i\sum_{J=1}^{N}\hat B_{J\mu}}\Big)=
\exp\Big({i\hat p_\mu -i\sum_{J=1}^{N-1}\hat B_{J\mu}-i\hat B_{N\mu}}\Big)=\\
&\exp\Big({i\hat p_\mu -i\sum_{J=1}^{N-1}\hat B_{J\mu}}\Big)
\exp\Big(-{i\beta_{N\mu}\hat B_{N\mu}}\Big)=\\
&e^{i\hat p_\mu }\exp\Big(-{i\sum_{J=1}^{N}\beta_{J\mu} \hat B_{J\mu}}\Big)=
e^{i\hat p_\mu }\prod_{J=1}^{N} e^{-i\beta_{J\mu} \hat B_{J\mu}}
\label{Z}\end{aligned} \end{equation}
where
\begin{equation} \begin{aligned}
\beta_{J\mu}=\frac{1-e^{-k_{J\mu}}}{k_{J\mu}}
\label{Z}\end{aligned} \end{equation}
Hence,
\begin{equation} \begin{aligned}
&\exp\big(i(\hat p_\mu- \sum_{J=1}^N A_{J \mu}e^{ {\bf k}_J i\partial_{\bf p}})\big)=
e^{i\hat p_\mu}\prod_{J=1}^{N} \exp\big(-i\beta_{J \mu} A_{J \mu}e^{ {\bf k}_J i\partial_{\bf p}} \big)=\\
&e^{i\hat p_\mu}\prod_{J=1}^{N} \sum_{n_J=0}^\infty\frac{(-i\beta_{J \mu} A_{J \mu})^{n_J}}{n_J!}e^{ n_J{\bf k}_J i\partial_{\bf p}}=\\
&e^{i\hat p_\mu}
\sum_{n_1=0}^\infty\frac{(-i\beta_{1\mu} A_{1\mu})^{n_1}}{n_1!}...
\sum_{n_N=0}^\infty\frac{(-i\beta_{N\mu} A_{N\mu})^{n_N}}{n_N!}
e^{(n_1{\bf k}_1+...+n_N{\bf k}_N) i\partial_{\bf p}}\\
\label{Z}\end{aligned} \end{equation}
For imaginary values of $\bf k$  we obtain

\begin{equation} \begin{aligned}
&\bra{{\bf p''}}
\exp\big(i(\hat p_\mu- \sum_{J=1}^N A_{J \mu}e^{ {\bf k}_J i\partial_{\bf p}})\big)\ket{\bf p'}=\\
&\bra{{\bf p''}}e^{i\hat p_\mu}
\sum_{n_1=0}^\infty\frac{(-i\beta_{1\mu} A_{1\mu})^{n_1}}{n_1!}...
\sum_{n_N=0}^\infty\frac{(-i\beta_{N\mu} A_{N\mu})^{n_N}}{n_N!}
e^{(n_1{\bf k}_1+...+n_N{\bf k}_N) i\partial_{\bf p}}
\ket{\bf p'}=\\
&\bra{{\bf p''}}e^{i\hat p_\mu}
\sum_{n_1=0}^\infty\frac{(-i\beta_{1\mu} A_{1\mu})^{n_1}}{n_1!}...
\sum_{n_N=0}^\infty\frac{(-i\beta_{N\mu} A_{N\mu})^{n_N}}{n_N!}
\ket{{\bf p}'-i(n_1{\bf k}_1+...+n_N{\bf k}_N)}=\\
&\bra{{\bf p''}}
\sum_{n_1=0}^\infty\frac{(-i\beta_{1\mu} A_{1\mu})^{n_1}}{n_1!}...
\sum_{n_N=0}^\infty\frac{(-i\beta_{N\mu} A_{N\mu})^{n_N}}{n_N!}e^{i p'_\mu+(n_1k_{1\mu}+...+n_Nk_{N\mu})}
\ket{{\bf p}'-i(n_1{\bf k}_1+...+n_N{\bf k}_N)}=\\
&\sum_{n_1=0}^\infty\frac{(-i\beta_{1\mu} A_{1\mu})^{n_1}}{n_1!}...
\sum_{n_N=0}^\infty\frac{(-i\beta_{N\mu} A_{N\mu})^{n_N}}{n_N!}e^{i p'_\mu+(n_1k_{1\mu}+...+n_Nk_{N\mu})}
\delta\Big[({\bf p}''-{\bf p}')+i(n_1{\bf k}_1+...+n_N{\bf k}_N)\Big]\\
\label{Z}\end{aligned} \end{equation}
The last expression allows to calculate the Weyl symbol of $F$:
\begin{equation} \begin{aligned}
&\Big[\exp\big(i(\hat p_\mu- \sum_{J=1}^N A_{J \mu}e^{ {\bf k}_J i\partial_{\bf p}})\big)\Big]_W=\\
&\int  d{\bf q} e^{i\bf qx}
\bra{{\bf p+q/2}}
\exp\big(i(\hat p_\mu- \sum_{J=1}^N A_{J \mu}e^{ {\bf k}_J i\partial_{\bf p}})\big)
 \ket{{\bf p-q/2}}=\\
&\int  d{\bf q} e^{i\bf qx}\\
&\sum_{n_1=0}^\infty\frac{(-i\beta_{1\mu} A_{1\mu})^{n_1}}{n_1!}...
\sum_{n_N=0}^\infty\frac{(-i\beta_{N\mu} A_{N\mu})^{n_N}}{n_N!}e^{i (p_\mu-q_\mu/2)+(n_1k_{1\mu}+...+n_Nk_{N\mu})}
\delta\Big[{\bf q}+i(n_1{\bf k}_1+...+n_N{\bf k}_N)\Big]=\\
&\sum_{n_1=0}^\infty\frac{(-i\beta_{1\mu} A_{1\mu})^{n_1}}{n_1!}...
\sum_{n_N=0}^\infty\frac{(-i\beta_{N\mu} A_{N\mu})^{n_N}}{n_N!} e^{(n_1{\bf k}_1+...+n_N{\bf k}_N)\bf x}e^{i p_\mu+\frac{1}{2}(n_1k_{1\mu}+...+n_Nk_{N\mu})}=\\
&\exp\Big[i\big(p_\mu-\sum_{J=1}^N \beta_{J\mu}e^{k_{J\mu}/2}A_{J \mu}e^{ {\bf k}_J {\bf x}}\big)\Big]
\label{Z}\end{aligned} \end{equation}
defining
\begin{equation} \begin{aligned}
a(k_{J\mu}) \equiv \beta_{J\mu}e^{k_{J\mu}/2}=
\frac{1-e^{-k_{J\mu}}}{k_{J\mu}}e^{k_{J\mu}/2}=
\frac{\sinh(k_{J\mu}/2)}{k_{J\mu}/2}
\label{Z}\end{aligned} \end{equation}
we come to
\begin{equation} \begin{aligned}
&\Big[\exp\big(i(\hat p_\mu- \sum_{J=1}^N A_{J \mu}e^{ {\bf k}_J i\partial_{\bf p}})\big)\Big]_W=\exp\Big[i\big(p_\mu-\sum_{J=1}^N a(k_{J\mu})A_{J \mu}e^{ {\bf k}_J {\bf x}}\big)\Big]
\label{Z}\end{aligned} \end{equation}
This representation has been derived for the imaginary values of $\bf k$. However, it may be extended to any complex values of $\bf k$  by analytical continuation.

\subsection  {Weyl symbol of $F=\exp\big[i(\hat p_\mu - A_{\mu}(i\partial_{\bf p}))\big]$ \\}

For the case of arbitrary function $A_{\mu}(i\partial_{\bf p})$  the Weyl symbol of $F=\exp\big[i(\hat p_\mu- A_{\mu}(i\partial_{\bf p}))\big]$ may be calculated using the above obtained expressions. We should represent $A$ in the form of the Laplace transformation:
\begin{equation} \begin{aligned}
A_{\mu}({\bf x})=\int \big( \tilde A_\mu({\bf k})e^{{\bf k}{\bf x}} +c.c. \big)d k
\label{Z}\end{aligned} \end{equation}
that is
\begin{equation} \begin{aligned}
A_{\mu}(i\partial_{\bf p})=\int \big( \tilde A_\mu({\bf k})e^{{\bf k}i\partial_{\bf p}} +c.c. \big)d k
\label{Z}\end{aligned} \end{equation}
In turn, this integral may be discretized and represented in the form of the series. As a result the Weyl symbol is given by
\begin{equation} \begin{aligned}
\Big[\exp\Big(i\hat p_\mu-i\int \big[\tilde A_\mu({\bf k})e^{{\bf k}i\partial_{\bf p}}+c.c. \big]dk\Big)\Big]_W=\exp\Big(ip_\mu-i{\cal A}_\mu({\bf x}) \Big)
\label{Z}\end{aligned} \end{equation}
where \revisionZ{
$${\cal A}_\mu({\bf x}) =  \int \big[a_\mu({\bf k})\tilde A_\mu({\bf k})e^{{\bf kx}}+c.c. \big]dk $$
and
\begin{equation} \begin{aligned}
a_\mu({\bf k}) =
\frac{1-e^{-k_{\mu}}}{k_{\mu}}e^{k_{\mu}/2}=
\frac{\sinh(k_{\mu}/2)}{k_{\mu}/2}
\label{Z}\end{aligned} \end{equation}}

\subsection{Weyl symbol of Wilson Dirac operator $\hat{Q}({\bf p}-A(i{\partial_{\bf p}}))$}

Now we are in the position to calculate the Weyl symbol of operator $\hat Q$ for the Wilson fermions in the presence of arbitrary electromagnetic field $A({\bf x})$. For the operator
\begin{equation} \begin{aligned}
\hat{Q}({\bf p}) =\sum_{k=1,2,3,4} \gamma^k g_k ({\bf p})-im({\bf p})
\label{WF}\end{aligned} \end{equation}
with
\begin{equation} \begin{aligned}
g_k({\bf p})=\sin( p_k) \quad\quad m({\bf p})=
m^{(0)}+\sum_{\nu=1}^4 (1-\cos(p_\nu))
\label{Z}\end{aligned} \end{equation}
we obtain
\begin{equation} \begin{aligned}
\Big[{Q}({\bf p}-A(i{\partial_{\bf p}}))\Big]_W =\sum_{k=1,2,3,4} \gamma^k {\rm sin} ( p_k - {\cal A}_k({\bf x}))-i (m^{(0)}+\sum_{\nu=1}^4 (1-\cos(p_\nu - {\cal A}_{\nu}({\bf x}))))
\label{WF}\end{aligned} \end{equation}
where $\cal A$ is the following transformation of electromagnetic field:
\begin{equation}
{\cal A}_\mu({\bf x}) =  \int \big[\frac{\sin(k_{\mu}/2)}{k_{\mu}/2}\tilde A_\mu({\bf k})e^{i{\bf kx}}+c.c. \big]dk \label{calA}
\end{equation}
while the original electromagnetic field had the form:
$${A}_\mu({\bf x}) =  \int \big[\tilde A_\mu({\bf k})e^{i{\bf kx}}+c.c. \big]dk $$

\section{Propagator of fermionic quasiparticles in the presence of external electromagnetic field}

\label{Sect5}

\subsection{Wigner transformation of the two point Green function}

Let us consider the cubic tight - binding model of Wilson fermions.
The two point Green function ${\cal G}(p_1,p_2)$ obeys equation
$$
{\hat Q}({\bf p}-A(i {\partial_{\bf p}}))){\cal G}({\bf p},{\bf q}) = \delta({\bf p} - {\bf q})
$$
The Wigner transformation of $\cal G$ is defined as
\begin{equation} \begin{aligned}
{\cal G}_W(x_n,p)\equiv\int dq e^{ix_nq} {\cal G}({p+q/2}, {p-q/2})\label{GWx}
\end{aligned} \end{equation}
It obeys the Groenewold equation
\begin{equation} \begin{aligned}
{\cal G}_W(x_n,p) \star Q_W(x_n,p) = 1 \label{Geq}
\end{aligned} \end{equation}

\subsection{Iterative solution of the Groenewold equation}

The Gronewold equation has the explicit form
\begin{equation}\begin{aligned}
&1={\cal G}_W(x_n,p) \star Q_W(x_n,p) =
{\cal G}_W(x_n,p)
e^{\frac{i}{2} \left( \overleftarrow{\partial}_{x_n}\overrightarrow{\partial_p}-\overleftarrow{\partial_p}\overrightarrow{\partial}_{x_n}\right )}
Q_W(x_n,p)
\label{GQW}\end{aligned}\end{equation}
Let us introduce the following notation
$$
\overleftrightarrow{\Delta} = \frac{i}{2} \left( \overleftarrow{\partial}_{x_n}\overrightarrow{\partial_p}-\overleftarrow{\partial_p}\overrightarrow{\partial}_{x_n}\right )
$$
Notice, that the action of this symbol on $Q_W(x_n,p) = Q(p-{\cal A}(x_n)) $ is given by
$$
Q^{-1}(p-{\cal A}(x_n)) \overleftrightarrow{\Delta}  Q(p-{\cal A}(x_n)) = - Q^{-1}(p-{\cal A}(x_n))  \frac{i}{2}  \overleftarrow{\partial}_{p_j} {\cal F}_{jk}(x_n)\overrightarrow{\partial_{p_k}}Q(p-{\cal A}(x_n))
$$
where ${\cal F}_{ij}$ is composed of the transformed gauge potential ${\cal A}$ of Eq. (\ref{calA}) in the way similar to the field strength:
$$
{\cal F}_{ij} = \partial_i {\cal A}_j - \partial_j {\cal A}_i
$$
Let us represent ${\cal G}_W(x_n,p)$ as the series
$$
{\cal G}_W(x_n,p) = {\cal G}^{(0)}_W(x_n,p) + {\cal G}^{(1)}_W(x_n,p) + {\cal G}^{(2)}_W(x_n,p)+...
$$
where the term ${\cal G}^{(i)}_W(x_n,p)$  has the order $i$ in powers of the derivatives of ${\cal A}$ with respect to $x_n$.
We come to the following series of equations
\begin{equation}\begin{aligned}
1=&{\cal G}^{(0)}_W(x_n,p)Q_W(x_n,p)\\
1=&{\cal G}^{(0)}_W(x_n,p)Q_W(x_n,p)  +{\cal G}^{(0)}_W(x_n,p) \overleftrightarrow{\Delta} Q_W(x_n,p)\\&+{\cal G}^{(1)}_W(x_n,p)Q_W(x_n,p)\\
1=&{\cal G}^{(0)}_W(x_n,p)Q_W(x_n,p)  +{\cal G}^{(0)}_W(x_n,p) \overleftrightarrow{\Delta} Q_W(x_n,p)+\frac{1}{2}{\cal G}^{(0)}_W(x_n,p) \overleftrightarrow{\Delta}^2 Q_W(x_n,p)\\&+{\cal G}^{(1)}_W(x_n,p)Q_W(x_n,p) +{\cal G}^{(1)}_W(x_n,p) \overleftrightarrow{\Delta} Q_W(x_n,p) \\&+{\cal G}^{(2)}_W(x_n,p)  Q_W(x_n,p)\\
& ... \\
1=& \sum_{i,k=0...n; \, i+k \le n} \frac{1}{k!} {\cal G}^{(i)}_W(x_n,p)  \overleftrightarrow{\Delta}^k
Q_W(x_n,p)
\label{GQWs}\end{aligned}\end{equation}
From this sequence we obtain
\begin{equation}\begin{aligned}
{\cal G}^{(0)}_W(x_n,p)=&Q^{-1}_W(x_n,p)\\
{\cal G}^{(1)}_W(x_n,p)=&-\Big[Q^{-1}_W(x_n,p)) \overleftrightarrow{\Delta} Q_W(x_n,p)\Big]Q^{-1}_W(x_n,p)\\
{\cal G}^{(2)}_W(x_n,p) =&-\Big[\frac{1}{2}{\cal G}^{(0)}_W(x_n,p) \overleftrightarrow{\Delta}^2 Q_W(x_n,p)+{\cal G}^{(1)}_W(x_n,p) \overleftrightarrow{\Delta} Q_W(x_n,p) \Big] Q^{-1}_W(x_n,p)\\
& ... \\
{\cal G}^{(n)}_W(x_n,p)=& - \sum_{k=0...n-1} \frac{1}{(n-k)!}
\Big[{\cal G}^{(k)}_W(x_n,p)  \overleftrightarrow{\Delta}^k
Q_W(x_n,p)\Big] Q^{-1}_W(x_n,p)
\label{GQWs}\end{aligned}\end{equation}
In the other words
\begin{equation}\begin{aligned}
{\cal G}^{(0)}_W(x_n,p)=&Q^{-1}_W(x_n,p)\\
{\cal G}^{(1)}_W(x_n,p)=&-\Big[Q^{-1}_W(x_n,p)) \overleftrightarrow{\Delta} Q_W(x_n,p)\Big]Q^{-1}_W(x_n,p)\\
{\cal G}^{(2)}_W(x_n,p) =&-\Big[\frac{1}{2}Q^{-1}_W(x_n,p)\overleftrightarrow{\Delta}^2 Q_W(x_n,p) -\Big[Q^{-1}_W(x_n,p)) \overleftrightarrow{\Delta} Q_W(x_n,p)\Big]Q^{-1}_W(x_n,p) \overleftrightarrow{\Delta} Q_W(x_n,p) \Big] Q^{-1}_W(x_n,p)\\
& ... \\
{\cal G}^{(n)}_W(x_n,p)=& - \sum_{k=0...n-1} \frac{1}{(n-k)!}
\Big[{\cal G}^{(k)}_W(x_n,p)  \overleftrightarrow{\Delta}^{n-k}
Q_W(x_n,p)\Big] Q^{-1}_W(x_n,p)
\label{GQWs}\end{aligned}\end{equation}

\subsection{The final form of the iterative solution}

We represent the results of this iterative solution as follows:
\revisionZ{\begin{equation}\begin{aligned}
{\cal G}_W(x_n,p)=& \sum_k {\cal G}^{(k)}_W(x_n,p)\\
{\cal G}^{(n)}_W(x_n,p) =&\sum_{\sum_{i=1...M} k_i = n} \, {\cal C}^{n}_{k_1 k_2 ... k_M} \, \Big[...\Big[Q^{-1}_W \overleftrightarrow{\Delta}^{k_1} Q_W\Big] Q^{-1}_W \overleftrightarrow{\Delta}^{k_2} Q_W\Big]Q^{-1}_W... \overleftrightarrow{\Delta}^{k_M} Q_W \Big] Q^{-1}_W
\label{GQWsf}\end{aligned}\end{equation}}
Let us substitute this expression to Eq. (\ref{GQWs}):
\begin{equation}\begin{aligned}
{\cal G}^{(n)}_W(x_n,p)=& - \sum_{m=0...n-1} \frac{1}{(n-m)!}\sum_{\sum_{i=1...M} k_i = m} \, {\cal C}^{m}_{k_1 k_2 ... k_M} \\& \Big[...\Big[Q^{-1}_W \overleftrightarrow{\Delta}^{k_1} Q_W\Big] Q^{-1}_W \overleftrightarrow{\Delta}^{k_2} Q_W\Big]Q^{-1}_W... \overleftrightarrow{\Delta}^{k_M} Q_W \Big] Q^{-1}_W\overleftrightarrow{\Delta}^{n-m}
Q_W \Big] Q^{-1}_W
\label{GQWsff}\end{aligned}\end{equation}
\revisionZ{Constants ${\cal C}^{n}_{k_1 k_2 ... k_M} $ satisfy the following equations, that determine them iteratively:
\begin{equation}\begin{aligned}
{\cal C}_0^0=1\\
{\cal C}^{k_1+...+k_M+k_{M+1}}_{k_1 k_2 ... k_Mk_{M+1}}=& - \frac{1}{k_{M+1}!} \, {\cal C}^{k_1+...+k_M}_{k_1 k_2 ... k_M}
\label{CC}\end{aligned}\end{equation}}
We obtain:
\begin{equation}\begin{aligned}
&{\cal C}_0^0=1\\
&{\cal C}^1_1 = -1\\
&{\cal C}^2_2 = -\frac{1}{2},\, {\cal C}^2_{1,1}=1\\
&{\cal C}^3_3 = -\frac{1}{3!}, \, {\cal C}^3_{12}=\frac{1}{2},\, {\cal C}^3_{2,1}=\frac{1}{2}, \, {\cal C}^3_{1,1,1}=-1\\
&{\cal C}^4_4 = -\frac{1}{4!}, \, {\cal C}^4_{13}=\frac{1}{3!},\, {\cal C}^4_{2,2}=\frac{1}{4}, \, {\cal C}^4_{1,1,2}=-1/2, \, {\cal C}^4_{3,1}=\frac{1}{3!}, \, {\cal C}^4_{1,2,1}=-\frac{1}{2}, \, {\cal C}^4_{2,1,1}=-\frac{1}{2}, \, {\cal C}^4_{1,1,1,1}=1\\
& ... \\
& {\cal C}^{k_1+...+k_M}_{k_1 k_2 ... k_M}= \frac{(-1)^M}{k_1! k_2! ... k_M!}
\\
& ...
\label{GQWr}\end{aligned}\end{equation}
We obtain the final form of the solution:
\revisionZZ{\begin{equation}\begin{aligned}
{\cal G}_W(x_n,p) =&Q^{-1}_W + \sum_{n=1...\infty}\sum_{\begin{array}{c} M=1...n\\k_1  +... + k_M = n\\ k_i\ne 0\end{array}} \, \frac{(-1)^M}{k_1! k_2! ... k_M!} \, \Big[...\Big[Q^{-1}_W \overleftrightarrow{\Delta}^{k_1} Q_W\Big] Q^{-1}_W \overleftrightarrow{\Delta}^{k_2} Q_W\Big]Q^{-1}_W... \overleftrightarrow{\Delta}^{k_M} Q_W \Big] Q^{-1}_W \\
=&\sum_{M=0...\infty}  \, \underbrace{\Big[...\Big[Q^{-1}_W(1- e^{\overleftrightarrow{\Delta}}) Q_W\Big] Q^{-1}_W (1-e^{\overleftrightarrow{\Delta}}) Q_W\Big]... (1-e^{\overleftrightarrow{\Delta}}) Q_W \Big]} Q^{-1}_W\\
& \hspace{5cm} \emph{M \,  brackets} \\
=&\sum_{M=0...\infty}  \,\underbrace{ \Big[...\Big[Q^{-1}_W(1- \star) Q_W\Big] Q^{-1}_W(1- \star) Q_W\Big]Q^{-1}_W... (1-\star) Q_W \Big]} Q^{-1}_W\\
& \hspace{5cm} \emph{M \,  brackets} \\
\label{GQWsff}\end{aligned}\end{equation}}
In the first row the sum may be extended to the values $M=n=0$, then the first term will be equal to $Q^{-1}_W$.
Let us introduce the product operator $\bullet$, which works as follows being combined with the star product introduced above:
$$
A \bullet B \star C = (AB) \star C, \quad A\star B \bullet C = (A \star B) \bullet C
$$
In the first equation $\star$  acts both on $AB$ and on $C$ while in the second equation it acts only on $A$ and $B$. These rules allow to write the above equation in the compact way:
\revisionZZ{\begin{equation}\begin{aligned}
{\cal G}_W(x_n,p) =&\sum_{M=0...\infty}  \,\underbrace{Q^{-1}_W(1- \star) Q_W\bullet Q^{-1}_W(1- \star) Q_W\bullet Q^{-1}_W... (1-\star) Q_W \bullet} Q^{-1}_W\\
& \hspace{5cm} {M \, \bullet -  products }\\
=&\sum_{M=0...\infty}  \,\Big(Q^{-1}_W(1- \star) Q_W\bullet\Big)^M  Q^{-1}_W
\label{GQWsffb}\end{aligned}\end{equation}}
We may  write symbolically:
\revisionZZ{\begin{equation}\begin{aligned}
{\cal G}_W(x_n,p) =&\Big(1- \,Q^{-1}_W(1- \star) Q_W\bullet\Big)^{-1}  Q^{-1}_W = \Big( \,Q^{-1}_W \star Q_W\bullet\Big)^{-1}  Q^{-1}_W
\label{GQWsffa}\end{aligned}\end{equation}}
The last expression serves also as the alternative proof of Eq. (\ref{GQWsff}) because we started from
${\cal G}_W\star Q_W =1$. We substitute Eq. (\ref{GQWsffa}) to the star product  ${\cal G}_W\star Q_W $ and obtain
\revisionZZ{\begin{equation}\begin{aligned}
{\cal G}_W\star Q_W =& \sum_{M=0...\infty}  \,\Big(Q^{-1}_W(1- \star) Q_W\bullet\Big)^M  Q^{-1}_W \star Q_W \\ & = -\sum_{M=0...\infty}  \,\Big(Q^{-1}_W(1- \star) Q_W\bullet\Big)^M  Q^{-1}_W (1-\star) Q_W + \sum_{M=0...\infty}  \,\Big(Q^{-1}_W(1- \star) Q_W\bullet\Big)^M\\ & = - \sum_{M=1...\infty}  \,\Big(Q^{-1}_W(1- \star) Q_W\bullet\Big)^M  + \sum_{M=0...\infty}  \,\Big(Q^{-1}_W(1- \star) Q_W\bullet\Big)^M   \\ & = \Big(Q^{-1}_W(1- \star) Q_W\bullet\Big)^0 = 1
\label{GQWsffa}\end{aligned}\end{equation}}

\section{Summary of obtained results and conclusions}

\label{Sect6}

In this paper we develop the Wigner Weyl formalism for the model of lattice Wilson fermions. The main results specified below are the explicit expression for the Weyl symbol of lattice Wilson Dirac operator and the explicit expression for the fermion propagator in the presence of {\it arbitrary} external electromagnetic field.  The main purpose of the work is to provide the necessary tools for the analytical studies of the lattice models. Such a study may precede in many cases the numerical simulations and thus may also improve the latter indirectly. The pure analytical investigation itself of the lattice models is relevant, for example, for the study of the so - called non - dissipative transport. (It is also called anomalous transport because it reveals the correspondence with chiral anomaly, scale anomaly, etc.)
In \cite{ZW1,ZW2,ZW3,ZW4,ZW5,ZW6} the two first terms (in the expansion in powers of the derivatives) of the present solution of the Groenewold equation were calculated, which give the linear response of vector/axial currents to the external field strength. It appears that in many cases the corresponding coefficients are the topological invariants in momentum space, i.e. they are not changed when the system is changed smoothly. This gives the efficient description of certain anomalous transport phenomena (chiral magnetic effect, chiral separation effect, anomalous quantum Hall effect, scale magnetic effect, etc).

Our present study  generalizes the approach of  \cite{ZW1,ZW2,ZW3,ZW4,ZW5,ZW6} essentially and gives the more powerful method of calculations. In  the present paper we represent the complete iterative solution of the Groenewold equation (to all orders of the derivative expansion) in the presence of arbitrarily varying external electromagnetic field. It will allow to investigate the anomalous transport in case of varying external field strength. We foresee, that the intimate relation between topology and the non - dissipative transport will survive here as well, but its description may, possibly, be governed by the more complicated topological invariants.

Let us repeat once again the obtained results.
The derived expression for the Weyl symbol of the lattice Wilson Dirac operator in the presence of arbitrary electromagnetic field appears to be surprisingly simple:
\begin{equation} \begin{aligned}
\Big[{Q}({\bf p}-A(i{\partial_{\bf p}}))\Big]_W =\sum_{k=1,2,3,4} \gamma^k {\rm sin} ( p_k - {\cal A}_k({\bf x}))-i (m^{(0)}+\sum_{\nu=1}^4 (1-\cos(p_\nu - {\cal A}_{\nu}({\bf x}))))
\label{WFs}\end{aligned} \end{equation}
where by $\cal A$ we denote the following transformation of the original electromagnetic field:
\begin{equation}
{\cal A}_\mu({\bf x}) =  \int \big[\frac{\sin(k_{\mu}/2)}{k_{\mu}/2}\tilde A_\mu({\bf k})e^{i{\bf kx}}+c.c. \big]dk \label{calAs}
\end{equation}
(The original electromagnetic field itself has the form:
${A}_\mu({\bf x}) =  \int \big[\tilde A_\mu({\bf k})e^{i{\bf kx}}+c.c. \big]dk $.)
Eq. (\ref{WFs}) is further used to calculate the Wilson fermion propagator in the presence of arbitrary electronagnetic field. Namely, we consider first the Wigner transformation $G_W$ of the propagator, and solve iteratively the Groenewold equation. This solution has the form:
\revisionZZ{\begin{equation}\begin{aligned}
{\cal G}_W(x_n,p) =&Q^{-1}_W + \sum_{n=1...\infty}\sum_{\begin{array}{c} M=1...n\\k_1  +... + k_M = n\\ k_i\ne 0\end{array}} \, \frac{(-1)^M}{k_1! k_2! ... k_M!} \, \Big[...\Big[Q^{-1}_W \overleftrightarrow{\Delta}^{k_1} Q_W\Big] Q^{-1}_W \overleftrightarrow{\Delta}^{k_2} Q_W\Big]Q^{-1}_W... \overleftrightarrow{\Delta}^{k_M} Q_W \Big] Q^{-1}_W \\
=&\sum_{M=0...\infty}  \, \underbrace{\Big[...\Big[Q^{-1}_W(1- e^{\overleftrightarrow{\Delta}}) Q_W\Big] Q^{-1}_W (1-e^{\overleftrightarrow{\Delta}}) Q_W\Big]... (1-e^{\overleftrightarrow{\Delta}}) Q_W \Big]} Q^{-1}_W \\
& \hspace{5cm} \emph{M \, { brackets}} \end{aligned}\end{equation}}
where
$
\overleftrightarrow{\Delta} = \frac{i}{2} \left( \overleftarrow{\partial}_{x_n}\overrightarrow{\partial_p}-\overleftarrow{\partial_p}\overrightarrow{\partial}_{x_n}\right )
$ .
The obtained solution for the Wigner transformation of the two point Green function allows to reconstruct the Green function itself. In momentum space we have:
\begin{equation} \begin{aligned}
{\cal G}({p+q/2}, {p-q/2}) = \frac{1}{(2\pi)^D}\sum_{n} e^{- i x_n q}  {\cal G}_W(x_n,p)
\end{aligned} \end{equation}
while in coordinate space the propagator is given by
\begin{equation} \begin{aligned}
\tilde{\cal G}(z_n, y_n) = \int \frac{dp}{(2\pi)^D} e^{ i (z_n - y_n) p}  {\cal G}_W\Big(\frac{z_n+y_n}{2},p\Big)
\end{aligned} \end{equation}

The presented scheme allows to calculate, in principle, the propagator of Wilson fermions on the background of {\it any} external electromagnetic field. It would be interesting to apply it, for example, to the calculation of the propagator in the presence of varying external magnetic field of the particular form or in the presence of the particular electromagnetic wave. It is also worth mentioning, that the above obtained expressions may be generalized relatively easily to the case, when the external gauge field is non - Abelian. Then, those expressions may be used for the calculation of various observables in QCD using numerical simulations. Many of those observables are expressed through the vacuum average of the products of fermion Green functions in the presence of dynamical gauge field. We should then first use the expression for the propagator in the presence of external gauge field. The product of such propagators is represented as a series in powers of the derivatives of $Q_W$. In quenched approximation each term should be averaged over the gauge field with the weight equal to ${\rm exp}(-S)$, where $S$ is the pure gauge field action. This is to be done using numerical simulations of the pure gauge theory. Depending on the particular problem several simplifications may be made: for example, only a few first terms in the series in powers of the derivatives of $Q_W$ may be taken. In order to take into account the dynamical fermions we should also include into the consideration the fermion Determinant using the hybrid  Monte -  Carlo (HMC) algorithm. Our expression for the propagator in the presence of external field may also become the source of a modification of the HMC algorithm. But this is to be the subject of a separate study.

Another interesting continuation of the presented research is the extension of the Wigner - Weyl formalism to the more complicated lattice models defined on the rectangular lattices to be used for the regularization of the continuum quantum field theory. It would be interesting, in particular, to extend the proposed formalism to the model with overlap fermions that are most suitable for the investigation of QCD in the chiral limit. We foresee certain difficulties in such an extension related to the calculation of the Weyl symbol of the overlap Dirac operator caused by the structure of the latter essentially different from that of the Wilson fermions.

The presented results also may be extended to the lattice models with the non - rectangular lattice, that will allow to calculate analytically the fermionic quasiparticle propagator in various tight - binding models of the solid state physics. Unlike the case of the overlap fermions, in this extension in many cases the general structure of operator $\hat Q$ remains similar to that of Eq. (\ref{WF0}) with certain new basis matrices $\gamma_k$ instead of the Dirac matrices and with the arguments of $g$ and $m$ that depend on the projection of momenta to vectors of reciprocal lattice.

To conclude, in the present paper we propose the extension of the Wigner - Weyl formalism to the lattice  quantum field theory with Wilson fermions. The proposed formalism gives the efficient algorithm for the calculation of the Wilson fermions propagator in the presence of {\it arbitrary} external electromagnetic field. This technique may be extended in the straightforward way to the wide range of the non - hypercubic lattice models relevant for the description of the solid state physics. Thus we expect that the proposed technique may have applications both to the quantum field theory in lattice regularization and to the solid state physics. Using the lattice Wilson fermion propagator in the presence of arbitrary external electromagnetic field this will be possible to investigate via numerical simulations, for example, various effects in quark matter in the presence  of in - homogeneous external electromagnetic field. On the solid state physics side the direct applications will become possible when the similar formalism will be developed for the crystal lattice of more complicated and more realistic form. Then our formalism will allow to investigate various phenomena typical for the interaction of electronic quasiparticles with photons.

Both authors kindly acknowledge useful discussions with Z.V.Khaidukov.

\section*{Appendix I. Fourier transform}
In this Appendix we accumulate for the convenience of the reader the well - known expressions for the Fourier transform and Fourier series widely used throughout the text of the paper. For simplicity we consider the one - dimensional constructions.

\subsection{Continuous coordinates - continuous momenta}

Fourier transform:
\begin{equation}\begin{aligned}
f(x)&=\frac{1}{\sqrt{2\pi}}\int_{-\infty}^\infty dk e^{ikx} \tilde{f} (k)\quad\quad\quad
\tilde{f}(k)&=\frac{1}{\sqrt{2\pi}}\int_{-\infty}^\infty dx e^{-ikx} f(x)
\label{Z}\end{aligned}\end{equation}
The integral representation for the delta functions
\begin{equation}\begin{aligned}
f(x)=&\frac{1}{2\pi}\int_{-\infty}^\infty dk e^{ikx} \int_{-\infty}^\infty dx' e^{-ikx'} f(x')= \int_{-\infty}^\infty dx' f(x') \frac{1}{2\pi}\int_{-\infty}^\infty dk e^{ik(x-x')}
\label{Z}\end{aligned}\end{equation}
hence
\begin{equation}\begin{aligned}
\delta(x-x')&=\frac{1}{2\pi}\int_{-\infty}^\infty e^{ ik(x-x')}dk
\end{aligned}\end{equation}
and
\begin{equation}\begin{aligned}
\tilde{f}(k)&=\int_{-\infty}^\infty dx e^{-ikx} \frac{1}{2\pi}\int_{-\infty}^\infty dk' e^{ik'x} \tilde{f} (k')=
\int_{-\infty}^\infty dk' \tilde{f} (k')  \frac{1}{2\pi} \int_{-\infty}^\infty dx e^{-ix(k-k')}
\label{Z}\end{aligned}\end{equation}
hence
\begin{equation}\begin{aligned}
\delta(k-k')&=\frac{1}{2\pi}\int_{-\infty}^\infty e^{- ix(k-k')}dx\\
\end{aligned}\end{equation}

\subsection{Compact coordinates, $x\in [0,L]$  - discrete momenta}
Fourier series:
\begin{equation}\begin{aligned}
f(x)&=\sum_{n=0}^\infty e^{ik_0nx} \tilde{f} (k_n)\quad\quad k_n=k_0n\quad k_0=\frac{2\pi}{L}\\
\tilde{f}(k_n)&=\frac{1}{L}\int_{0}^L dx e^{-ik_nx} f(x)
\label{Z}\end{aligned}\end{equation}
The representation of the delta functions through the integrals/series:
\begin{equation}\begin{aligned}
f(x)&=\sum_{n=0}^\infty e^{ik_0nx} \frac{1}{L}\int_{0}^L dx' e^{-ik_0nx'} f(x')=\int_{0}^L dx' f(x') \frac{1}{L} \sum_{n=0}^\infty e^{ik_0n(x-x')}
\label{Z}\end{aligned}\end{equation}
hence
\begin{equation}\begin{aligned}
\delta(x-x')=\frac{1}{L} \sum_{n=0}^\infty e^{ik_0n(x-x')}
\label{Z}\end{aligned}\end{equation}
and
\begin{equation}\begin{aligned}
\tilde{f}(k_n)=\frac{1}{L}\int_{0}^L dx e^{-ik_0nx} \sum_{m=0}^\infty e^{ik_0mx} \tilde{f} (k_m)=\sum_{m=0}^\infty \tilde{f} (k_m) \frac{1}{L}\int_{0}^L dx e^{-ik_0x(n-m)}
\label{Z}\end{aligned}\end{equation}
hence
\begin{equation}\begin{aligned}
\delta(k_n-k_m)=\frac{1}{L}\int_{0}^L dx e^{-ik_0x(n-m)}
\label{Z}\end{aligned}\end{equation}
\subsection{Discrete coordinates - the compact region of momenta, $k\in [0,\frac{2\pi}{L}]$}
Fourier series:
\begin{equation}\begin{aligned}
\tilde{f}(k)&=\sum_{n=0}^\infty e^{-ikx_n} f(x_n) \quad\quad\quad x_n=nL\\
f(x_n)&=\frac{1}{2\pi/L}\int_{0}^{2\pi/L} dk e^{ikx_n} \tilde{f}(k)
\label{Z}\end{aligned}\end{equation}
The representation of the delta functions as the integrals/series:
\begin{equation}\begin{aligned}
\tilde{f}(k)=\sum_{n=0}^\infty e^{-iknL} \frac{L}{2\pi}\int_{0}^{2\pi/L} dk' e^{ik'nL} \tilde{f}(k')=\int_{0}^{2\pi/L} dk' \tilde{f}(k') \frac{L}{2\pi} \sum_{n=0}^\infty e^{-inL(k-k')}
\label{Z}\end{aligned}\end{equation}
hence
\begin{equation}\begin{aligned}
\delta(k-k')= \frac{L}{2\pi} \sum_{n=0}^\infty e^{-inL(k-k')}
\label{Z}\end{aligned}\end{equation}
and
\begin{equation}\begin{aligned}
f(x_n)&=\frac{L}{2\pi}\int_{0}^{2\pi/L} dk e^{iknL} \sum_{m=0}^\infty e^{-ikmL} f(x_m) =
\sum_{m=0}^\infty f(x_m) \frac{L}{2\pi}\int_{0}^{2\pi/L} dk e^{ikL(n-m)}
\label{Z}\end{aligned}\end{equation}
hence
\begin{equation}\begin{aligned}
\delta(x_n-x_m)= \frac{L}{2\pi}\int_{0}^{2\pi/L} dk e^{ikL(n-m)}
\label{Z}\end{aligned}\end{equation}

\section*{Appendix II. An expression for the P ordered exponent. }

The path-ordering operator is defined as follows
\begin{equation}
P(f(x)g(y))\equiv \theta(x-y)g(y)f(x)+\theta(y-x)f(x)g(y)
\label{1} \end{equation}
This definition is consistent with the following equation
\begin{equation}
P\,e^{\int_{x_1}^{x_2}\,dx \,f(x)}\equiv \lim_{\Delta \to 0} \prod_{n=0}^N e^{\Delta f(x_1+n\Delta)}
\label{1} \end{equation}
where $N=\frac{x_2-x_1}{\Delta}$.\\
\\
\begin{center}{\bf Lemma}\end{center}

Let $\hat{B}$ and $\hat{C}$ be operators. Then
\begin{equation}
e^{\hat{B}+\hat{C}}=Pe^{\int_{0}^{1}due^{\hat{B}u}\hat{C}e^{-\hat{B}u}}e^{\hat{B}}
\label{Lemma1} \end{equation}

\begin{center}{\it Proof}\end{center}

After the discretization we come to
\begin{equation}
I=Pe^{\int_{0}^{1}due^{\hat{B}u}\hat{C}e^{-\hat{B}u}}e^{\hat{B}}=\lim_{\Delta \to 0}\left[\prod_{n=0}^{N=\frac{1}{\Delta}} e^{\Delta e^{n\Delta\hat{B}}\hat{C}e^{-n\Delta\hat{B}}}\right]e^{\hat{B}}
\label{1} \end{equation}
that is
\begin{equation}\begin{aligned}
I=\left[ e^{\Delta e^{0\Delta\hat{B}}\hat{C}e^{-0\Delta\hat{B}}} \right]
   \left[ e^{\Delta e^{1\Delta\hat{B}}\hat{C}e^{-1\Delta\hat{B}}} \right]...
   \left[ e^{\Delta e^{(1-\Delta)\hat{B}}\hat{C}e^{-(1-\Delta)\hat{B}}} \right]
   \left[ e^{\Delta e^{\hat{B}}\hat{C}e^{-\hat{B}}} \right]
e^{\hat{B}}
\label{1} \end{aligned}\end{equation}\\
Next, using relation  $e^{\hat{O}\hat{A}\hat{O}^{-1}}=1+\hat{O}\hat{A}\hat{O}^{-1}+\frac{1}{2!}\hat{O}\hat{A}\hat{O}^{-1}\hat{O}\hat{A}\hat{O}^{-1}+...=\hat{O}e^{\hat A}\hat{O}^{-1}$, we may rewrite
$I$  as follows
\begin{equation}\begin{aligned}
I=&\left[ e^{\Delta\hat{C}} \right]
   \left[ e^{\Delta\hat{B}}e^{\Delta\hat{C}}e^{-\Delta\hat{B}} \right]
   \left[ e^{\Delta\hat{2B}}e^{\Delta\hat{C}}e^{-\Delta\hat{2B}} \right]...\\
  & \left[ e^{(1-\Delta)\hat{B}}e^{\Delta\hat{C}}e^{-(1-\Delta)\hat{B}} \right]
   \left[ e^{\hat{B}}e^{\Delta\hat{C}}e^{-\hat{B}} \right]
e^{\hat{B}}=e^{\Delta\hat{C}}\left[e^{\Delta\hat{B}}e^{\Delta\hat{C}}\right]^N\approx\\
&(1+\Delta \hat{C})\left[(1+\Delta(\hat{B}+\hat{C})+\Delta^2\hat{B}\hat{C})\right]^N\approx\\
&\left[(1+\Delta(\hat{B}+\hat{C})\right]^N=e^{\hat{B}+\hat{C}}
\label{1} \end{aligned}\end{equation}

Let us also prove the following

\begin{center}{\bf Lemma}\end{center}
\begin{equation}
\mathrm{P}\exp\left[i\int_{u_1}^{u_2} e^{ipu} A(i\partial_p) e^{-ipu} du\right]e^{-ipx}=\exp \left[ {i \int\limits_{0}^{u_2-u_1}A(x+u)du-ipx} \right]
\label{Lemma2} \end{equation}
\begin{center}{\it Proof}\end{center}

\begin{equation}\begin{aligned}
&\mathrm{P}\exp\left[i\int_{u_1}^{u_2} e^{ipu} A(i\partial_p) e^{-ipu} \right]e^{-ipx}=
\left[
\lim_{\Delta \to 0}\prod_{n=0}^{N=\frac{u_2-u_1}{\Delta}}
e^{i\Delta e^{ipn\Delta} A(i\partial_p)e^{-ipn\Delta}}
\right]e^{-ipx}=\\
&\left[
\lim_{\Delta \to 0}\prod_{n=0}^{N=\frac{u_2-u_1}{\Delta}}
e^{ipn\Delta} e^{i\Delta A(i\partial_p)}e^{-ipn\Delta}
\right]e^{-ipx}=\\
&\left[
\lim_{\Delta \to 0}
\left(e^{i\Delta A(i\partial_p)}\right)
\left(e^{ip\Delta} e^{i\Delta A(i\partial_p)}e^{-ip\Delta}\right)...
\left(e^{ipN\Delta} e^{i\Delta A(i\partial_p)}e^{-ipN\Delta}\right)
\right]e^{-ipx}=\\
&\left[
\lim_{\Delta \to 0}
\left(e^{i\Delta A(i\partial_p)}\right)
\left(e^{ip\Delta} e^{i\Delta A(i\partial_p)}\right)...
\left(e^{ip\Delta} e^{i\Delta A(i\partial_p)}\right)e^{-ip(u_2-u_1)}
\right]e^{-ipx}=\\
&\lim_{\Delta \to 0}
\left(e^{i\Delta A(i\partial_p)}\right)
\left(e^{ip\Delta} e^{i\Delta A(i\partial_p)}\right)^Ne^{-ip(x+(u_2-u_1))}=\\
&\lim_{\Delta \to 0}
\left(e^{i\Delta A(i\partial_p)}\right)
\left(e^{ip\Delta} e^{i\Delta A(i\partial_p)}\right)^{N-1}e^{ip\Delta} e^{i\Delta A(x+(u_2-u_1))}e^{-ip(x+(u_2-u_1))}=\\
&\lim_{\Delta \to 0}
\left(e^{i\Delta A(i\partial_p)}\right)
\left(e^{ip\Delta} e^{i\Delta A(i\partial_p)}\right)^{N-1}e^{-ip(x+(u_2-u_1)-\Delta)}e^{i\Delta A(x+(u_2-u_1))}=\\
&\lim_{\Delta \to 0}
\left(e^{i\Delta A(i\partial_p)}\right)
\left(e^{ip\Delta} e^{i\Delta A(i\partial_p)}\right)^{N-2}e^{-ip(x+(u_2-u_1)-2\Delta)}e^{i\Delta \left(A(x+(u_2-u_1)-\Delta)\right)+A(x+(u_2-u_1))}=\\
&\lim_{\Delta \to 0}
\exp\left({i\sum_{n=0}^{N=\frac{u_2-u_1}{\Delta}}\Delta A(x+n\Delta)}\right)
e^{-ipx}=
\exp\left(i\int\limits_{0}^{u_2-u_1}duA(x+u)-ipx\right)
\end{aligned}\end{equation}

\section*{Appendix III. BCH formula for the particular  Lie algebra}

In this section we derive the particular case of the BCH formula, that corresponds to the Lie algebra of operators composed of the basis elements $X$, $Y$ with the commutation relation $[X,Y]=\alpha Y$.
We are going to prove the following
\begin{center}{\bf Lemma}\end{center}
\begin{equation} \begin{aligned}
e^{X}e^{Y}=\exp(X+\frac{\alpha}{1-e^{-\alpha}}Y)
\label{Z}\end{aligned} \end{equation}
or, alternatively
\begin{equation} \begin{aligned}
\exp(X+Y) =e^{X}e^{\beta Y}
\label{Z}\end{aligned} \end{equation}
where $\beta=\frac{1-e^{-\alpha}}{\alpha}$.
\begin{center}{\it Proof}\end{center}

Let us define
\begin{equation} \begin{aligned}
A(t)=\exp(X+tY), \, B(t)= e^{X}e^{\beta tY}
\label{Z}\end{aligned} \end{equation}
The  Lemma will be proved if we will show, that $A(t)$ and $B(t)$ obey the same first order differential equation, while $A(0)=B(0)$. The latter requirement is obvious. Let us calculate the derivatives:
\begin{equation} \begin{aligned}
\partial_t A(t)&=\partial_t \exp(X+tY)\\
&=\partial_t \exp\left(\frac{X+tY}{N}\right)^N\\
&=\sum_{k=0}^{N-1}\exp\left(\frac{X+tY}{N}\right)^k \left[\partial_t \exp\left(\frac{X+tY}{N}\right) \right] \exp\left(\frac{X+tY}{N}\right)^{N-k-1}
\label{Z}\end{aligned} \end{equation}
taking the limit $N\to\infty$, one gets
\begin{equation} \begin{aligned}
\partial_t \exp\left(\frac{X+tY}{N}\right)= \frac{Y}{N}
\label{Z}\end{aligned} \end{equation}
and the sum in this case becomes an integral
\begin{equation} \begin{aligned}
\partial_t A(t)&=\int_{0}^{1} e^{u(X+tY)} Y e^{(1-u)(X+tY)}du
\label{Z}\end{aligned} \end{equation}
where $u=\frac{k}{N}$.

In the following we will use the identity
\begin{equation} \begin{aligned}
e^{u(X+tY)} Y =Y e^{u(X+\alpha+tY)}
\label{id3}\end{aligned} \end{equation}
that follows from the obvious relation $u(X+tY)Y=Yu(X+\alpha +tY)$ based on the commutation relation of $X$ and $Y$. Hence,
\begin{equation} \begin{aligned}
\partial_t A(t)&=Ye^{(X+tY)}\int_{0}^{1} e^{u\alpha}du=YA(t)\frac{e^\alpha-1}{\alpha}
\label{Z}\end{aligned} \end{equation}
On the other hand, the derivative of $B(t)$ is given by
\begin{equation} \begin{aligned}
\partial_t B(t)=\partial_t  e^{X}e^{\beta tY}=e^{X}\beta Y e^{\beta tY}=Ye^{X} e^{\beta tY} \beta e^\alpha=YB(t)\frac{e^\alpha-1}{\alpha}
\label{Z}\end{aligned} \end{equation}
(Here we used Eq. (\ref{id3}) for $u=1,t=0$.) One can see, that indeed $A(t)$ and $B(t)$ obey the same first order differential equation. There is only one solution of this equation that obeys the  initial condition $B(0)=A(0)$, and this proves the Lemma.

\end{document}